\documentclass[
 aps,
 prl,
 amsmath,amssymb,
 reprint,
]{revtex4-1}

\usepackage[dvips]{graphicx}
\usepackage{dcolumn}
\usepackage{bm}
\usepackage[]{hyperref}
\hypersetup{
    colorlinks=true, 
    linkcolor=blue,
    citecolor=blue,
    filecolor=blue,
    urlcolor=blue,
    }
\usepackage[all]{hypcap}

\begin{document}
\title[]{Long Distance Spin Transport in High Mobility Graphene on Hexagonal Boron Nitride}

\author{P. J. Zomer}
\email{p.j.zomer@rug.nl}

\author{M. H. D. Guimar\~{a}es}

\author{N. Tombros}

\author{B. J. van Wees}

\affiliation{Physics of Nanodevices, Zernike Institute for Advanced Materials, University of Groningen, Groningen, The Netherlands
}

\date{\today}

\begin{abstract}

We performed spin transport measurements on boron nitride based single layer graphene devices with mobilities up to 40~000~cm${^2}$V$^{-1}$s$^{-1}$.
We could observe spin transport over lengths up to 20~$\mu$m at room temperature, the largest distance measured so far for graphene.
Due to enhanced charge carrier diffusion, spin relaxation lengths are measured up to 4.5~$\mu$m. 
The relaxation times are similar to values for lower quality SiO$_2$ based devices, around 200~ps.
We find that the relaxation rate is determined in almost equal measures by the Elliott-Yafet and D'Yakonov-Perel mechanisms.

\end{abstract}



\maketitle

The potential of graphene~\cite{RiseofG} as an emerging material for spintronics has been established, revealing spin relaxation lengths $\lambda$ of 2~$\mu$m at room temperature~\cite{TombrosSpin}.
Spins relax over a length $\lambda=\sqrt{D_s\tau_s}$, where D$_s$ is the spin diffusion constant and $\tau_s$ the spin relaxation time. One straightforward way to achieve spin transport over larger distances is to enhance D$_s$ by fabricating high mobility devices. On the other hand, $\tau_s$ is theoretically predicted to range up to hundreds of nanoseconds~\cite{HuertaSO}.
However, observations made in the recent years by experimentalists~\cite{TombrosSpin,JozsaLinScale,PopinciucR,HanReview,HanTunnelCts,HanMuDep,JoSpinRelaxLinDisp,GuimaraesSuspSpin,AvsarCVD,MaassenSiC,MaassenLocStatesSiC} do not match up to the high expectations set by theory. Measurements typically indicate $\tau_s$ to be in the hundred picoseconds range and the discrepancy between theory and experiment and the exact relaxation mechanism remain yet unclear. Some works suggest that spin relaxation is dominated by the Elliott-Yafet (EY) mechanism~\cite{OchoaEY,AvsarCVD,JozsaLinScale}, where $\tau_s$ is proportional to the momentum relaxation time $\tau_p$ and spins lose their information during scattering events. Other efforts indicate that the D'Yakonov-Perel (DP) mechanism is stronger~\cite{HuertaSO,ZhangRandomRashba,ErtlerRoleOfSubstr}, where $\tau_s$ is inversely proportional to $\tau_p$ and spins dephase in between scattering events. 

In identifying the limiting factors on spin transport in graphene, the substrate deserves special attention. 
For charge transport it has already been shown that the standard silicon oxide substrate reduces the mobility of charge carriers considerably~\cite{MartinSiO2}. The SiO$_2$ substrate is expected to also affect the spin relaxation in graphene through its roughness, trapped charges and surface phonons~\cite{ErtlerRoleOfSubstr}. 
One approach to reduce the substrate roughness and screen impurities is to use epitaxial graphene on silicon carbide~\cite{MaassenSiC,DlubakSiC}. However, the presence of localized states is believed to affect spin transport in this system~\cite{MaassenLocStatesSiC}.
Alternatively, eliminating the influence of the substrate by suspending the graphene flake yields a 3 orders of magnitude increase in mobility~\cite{BolotinSuspended,TombrosSuspendedPolymer}. 
Suspended spintronic graphene devices have been studied and a lower bound for $\tau_s$ of $\sim$200~ps was found~\cite{GuimaraesSuspSpin}. Determination of the actual value was however not possible since the presence of local supports for the suspended device was found to dominate the extraction of $\tau_s$. 

Atomically flat hexagonal boron nitride (h-BN) was found to be a much better substrate than SiO$_2$ for high quality graphene electronics~\cite{DeanBN,XueSTM,ZomerTransfer}, yielding a 2 orders of magnitude increase in mobility. In this manuscript we present spin transport measurements on h-BN based graphene devices, which give access to a higher mobility regime than explored so far in graphene spin transport, while overcoming SiO$_2$ related issues such as the roughness and the presence of trapped charges.

Devices are made by transfer of a graphene flake (HOPG grade ZYA) onto a h-BN crystal, typically $\sim$20~nm thick, following a transfer recipe described in detail elsewhere~\cite{ZomerTransfer}. The h-BN crystals are mechanically cleaved from a commercially available boron nitride powder (Momentive PolarTherm, Grade PT110). Electrical contacts to the graphene flake are made using standard EBL techniques. We first deposit either aluminum or titanium, in both cases in two steps of 0.4~nm, with an oxidation step after each deposition. Secondly we deposit 65~nm cobalt in order to have spin sensitive contacts. The oxide barrier at the contact interface is required to tackle the conductivity mismatch problem, i.e. to prevent spin relaxation through the contacts. Typically the contact resistance is in the order of $\sim$10~k$\Omega$. The standard recipe for high quality graphene devices on h-BN requires a final anneal step in Ar/H$_2$ flow at 330~$^{\circ}$C for 8~hours. We found that this treatment degrades our Co contacts so that they lose their spin injection and detection properties. Therefore we omitted this step here. Also no other annealing steps have been used, keeping the fabrication process the same as for SiO$_2$ based devices. Measurements are all done in vacuum ($\sim1\times10^{-7}$~mbar), using standard AC lock-in techniques with currents up to 5~$\mu$A.
 
\begin{figure}[]
\includegraphics[width=55mm]{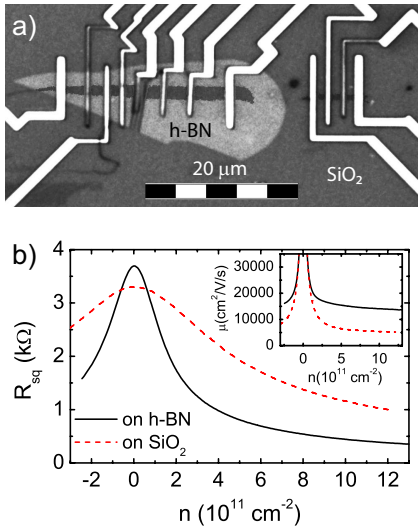}
\caption{
\label{fig:Fig1} (Color online) 
(a) Optical micrograph of device 2, partly on SiO$_2$ (right side) and partly on h-BN (left side).
(b) Square resistance versus charge carrier density measured on both a SiO$_2$ and a h-BN based parts of the same graphene flake. The inset shows the respective field effect mobilities.
}
\end{figure} 

To characterize our devices we determine the field effect mobility $\mu$ of the charge carriers: $\mu = 1/neR_{sq}$. Here $e$ is the electron charge and $n$ is the charge carrier density. The latter is calculated using $n=\frac{C_g}{e}(V-V_{D})$, where $V$ is the applied gate voltage, $V_D$ is gate voltage at which the charge neutrality point is found and $C_g\approx67$~$\mu$F m$^{-2}$ is the geometric gate capacitance for 500~nm SiO$_2$ and 20~nm h-BN. For device 2 shown in Fig.\ref{fig:Fig1}a we deposited the graphene flake partly on top of h-BN and partly on SiO$_2$, allowing for a direct comparison between charge carrier transport for both cases. The obtained values for $R_{sq}$ are presented in Fig.\ref{fig:Fig1}b, the inset shows the calculated mobility. The quality improvement due to the h-BN is directly reflected by the enhanced mobility. The inflection point mobilities at room temperature for the three devices presented here are for 1: 40~000~$cm^{2}V^{-1}s^{-1}$, 2: 21~000~$cm^{2}V^{-1}s^{-1}$ and 3: 14~000~$cm^{2}V^{-1}s^{-1}$. The reduction in full width at half maximum of the Dirac peak indicates smaller fluctuations in charge carrier density for the h-BN supported graphene flake~\cite{BolotinSuspended}. We find that despite omitting cleaning steps, the charge transport quality is well above that of SiO$_2$ based graphene spin transport devices.

\begin{figure}[b]
\includegraphics[width=85mm]{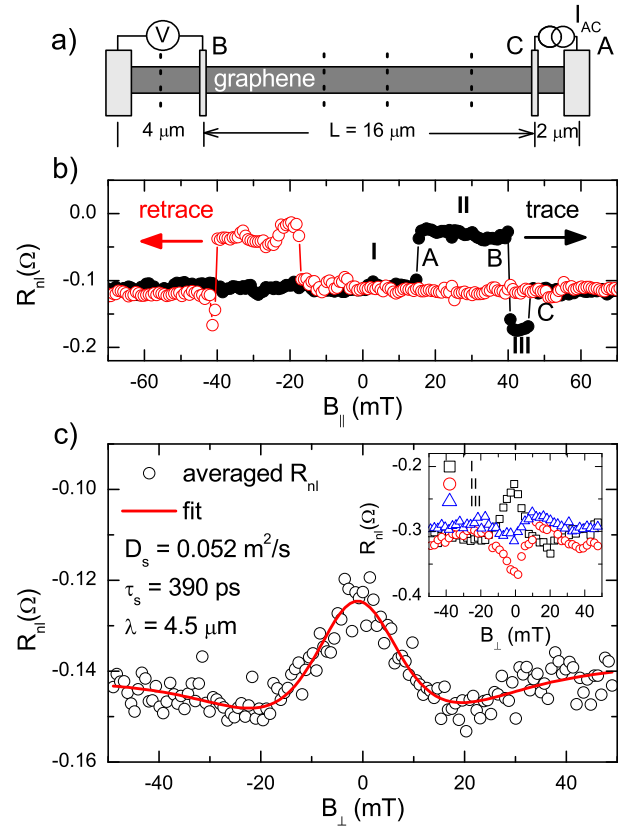}
\caption{
\label{fig:Fig2} (Color online) 
(a) Schematic showing the non-local 4 terminal geometry for device 3. Contacts not used in the measurement are represented by dashed lines.
(b) Spin valve measurement showing the non-local resistance versus magnetic field, parallel to the electrodes, at $n = 1.5\times10^{12}$~$cm^{-2}$. Switching of electrodes A to C shows up in the measurement.
(c) Hanle spin precession measurement ($R_{nl}$ versus perpendicular magnetic field) averaged between measurements at levels I to III (original data in inset). The solid line shows the fit from which $\tau_s$ and $D_s$ are extracted.
}
\end{figure}  

\begin{figure}[]
\includegraphics[width=85mm]{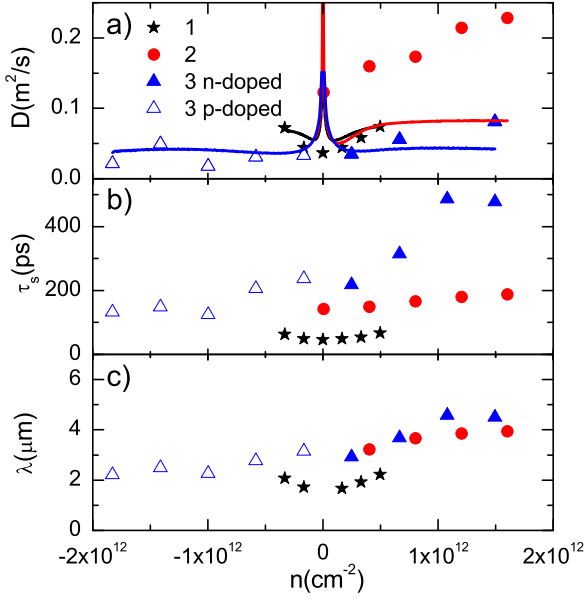}
\caption{
\label{fig:Fig3} (Color online) 
Room temperature data extracted from precession measurements for device 1, 2 and 3 as function of charge carrier density, with a respective injector detector spacing of 3.5, 3.5 and 2~$\mu$m. The barriers are made of Al$_2$O$_3$ for device 1 and 2 and of TiO$_2$ for device 3. (a) Spin- and charge diffusion constants, symbols and lines respectively. The latter are extracted from charge transport.
(b) Spin relaxation times.
(c) Spin relaxation lengths.
}
\end{figure} 

For spin transport measurements we employ the 4 terminal non-local technique~\cite{Jedema}, schematically shown in Fig.\ref{fig:Fig2}a. 
We inject a spin polarized current to the graphene flake by sending an electrical current through the pair of contacts on the right side. The injected spins diffuse through the graphene and arrive at the detection circuit on the left. 
An external magnetic field applied parallel to the contacts allows for control over their magnetization. Variation in contact width between 130 and 800~nm ensures different coercivity. Spin valve measurements are taken by sweeping the parallel magnetic field while recording the non-local resistance $R_{nl}$, as shown in Fig.\ref{fig:Fig2}b. The switches that occur in one sweep can be traced back to switching the magnetization of a specific contact. What makes this room temperature spin valve measurement particularly interesting is the large contact spacing. The total length covered is 18~$\mu$m, with multiple switches showing up. Note that the largest length over which a spin signal has been observed is 20~$\mu$m for this device, which is the largest contact spacing that was available. This is a direct indication of an improved spin relaxation length. Note that the presence of several other electrodes (dashed lines in Fig.\ref{fig:Fig2}a) between the injection and detection circuits does not introduce considerable additional spin scattering. 

\begin{figure}[]
\includegraphics[width=85mm]{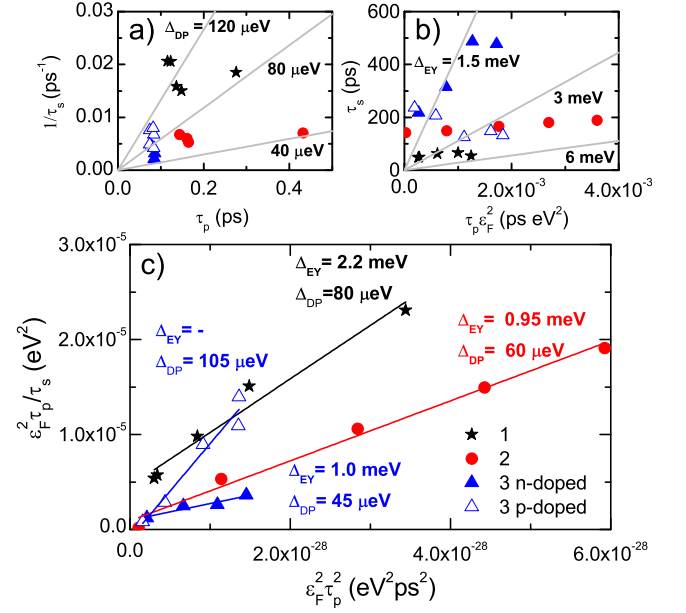}
\caption{
\label{fig:Fig4} (Color online) 
Room temperature data obtained from devices 1, 2 and 3. Linear relations allow for extraction of the spin-orbit coupling assuming (a) the DP mechanism or (b) the EY mechanism. The solid lines reflect the theoretical relation for several spin-orbit couplings. For both mechanisms the data deviates from theoretical expectations. (c) The combination of both the DP and EY mechanism allows for extraction of the respective spin-orbit couplings using Eq.\ref{eq:1}. The solid lines are linear fits.
}
\end{figure} 

We measure Hanle spin precession by applying an external magnetic field perpendicular to the graphene plane. The diffusing spins will precess at the Larmor frequency $\omega_{L}=g\mu_{B}B/\hbar$, where $g$ is the g-factor, $\mu_B$ is the Bohr magneton and $\hbar$ is the reduced Planck constant. Fig.\ref{fig:Fig2}c shows one set of precession data. Since in this case contact A contributes strongly to the spin signal, we measure precession for each magnetization geometry (at levels I to III), shown in the inset of Fig.\ref{fig:Fig2}c. Taking all three precession curves into account we can eliminate the contribution of the outer contact. The precession is fit using the one-dimensional Bloch equation in the steady state regime: $D_{s}\nabla^{2}\vec{\mu}_{s}-\frac{\vec{\mu}_{s}}{\tau_{s}}+\gamma\vec{B}\times\vec{\mu}_{s}=0$. Here $\vec{\mu}_{s}$ is the spin accumulation, $D_s$ is the spin diffusion constant and $\gamma$ is the gyromagnetic ratio. From a fit we acquire $D_s$ and $\tau_s$ and  hence we can calculate the spin relaxation length $\lambda=\sqrt{D_{s}\tau_{s}}$. 
For this particular dataset we find an increased $D_s = 0.052$~$m^2/s$ compared to SiO$_2$, with values $\sim0.02$~$m^2/s$. Interestingly, $\tau_s = 390$~$ps$ is not much different from a SiO$_2$ based device. The fact that it is higher than the typically observed 200~ps is due to the use of TiO$_2$ barriers instead of Al$_2$O$_3$, resulting in a reduced barrier roughness~\cite{HanTunnelCts}.

For devices 1 to 3 the parameters obtained by fitting room temperature precession measurements are shown in Fig.\ref{fig:Fig3}a to c. Values for $\tau_s$ are found in the range from 50 to 480~ps for various charge carrier densities, which is similar to what is found for lower mobility devices on SiO$_2$. For devices 2 and 3 D$_s$ does not match well with the charge diffusion constant D$_c$ (solid lines in Fig.\ref{fig:Fig3}a), which is obtained using the Einstein relation and charge transport measurements~\cite{JozsaLinScale}. Because our spin relaxation length is much larger than the contact spacing the determination of D$_s$ is less accurate and therefore we use D$_c$ to calculate $\lambda$. This way we obtain relaxation lengths up to 4.6~$\mu$m. We also measured spin transport at temperatures down to 4.2~K, which only led to a minor increase in $D_s$ and $\tau_s$. The behavior we measure for the p-doped part of device 3 deviates in the sense that $\tau_s$ decreases with increasing hole density. The cause for this is unclear.

In order to investigate the underlying spin relaxation mechanism in relation to the substrate and device quality, we can analyze the data from Fig.\ref{fig:Fig3} in the light of the DP and EY mechanisms by looking at the relation between $\tau_s$ and $\tau_p$. 
We extract the latter from charge transport measurements, using $\tau_{p}=\frac{2D_{c}}{v^{2}_{F}}$ where $v_F=10^{6}$~m/s is the Fermi velocity. 
For DP the relation between the spin relaxation time $\tau_s$ and momentum relaxation time $\tau_p$ is given by $\frac{1}{\tau_{s,DP}}=\left(\frac{4\Delta^{2}_{DP}}{\hbar^{2}}\right)\tau_{p}$, with $\Delta_{DP}$ as the effective spin-orbit coupling~\cite{ZhangRandomRashba}. 
In Fig.\ref{fig:Fig4}a this relation is plotted for three different values of $\Delta_{DP}$, together with the experimentally obtained data.
We observe that the linear trends described by theory are not reflected by our data.
On the other hand we can consider an EY mechanism, in which case the relation is given by $\tau_{s,EY}=\frac{\varepsilon^{2}_{F}\tau_p}{\Delta^{2}_{EY}}$, where $\varepsilon_{F}$ is the Fermi energy and $\Delta_{EY}$ is the spin-orbit coupling~\cite{HuertaSO,OchoaEY}. 
Fig.\ref{fig:Fig4}b shows this relation for three values of $\Delta_{EY}$. 
In this case the experimental data does show linear trends, but none of the sets intersect with the origin.
This cannot be attributed to broadening of the density of states or finite conductivity~\cite{JoSpinRelaxLinDisp}. 

Alternatively we can consider both mechanisms simultaneously, with the overall scattering rate given by $\frac{1}{\tau_{s}}=\frac{1}{\tau_{s,EY}}+\frac{1}{\tau_{s,DP}}$, which leads to the following relation: 
\begin{equation}
\label{eq:1}
\frac{\varepsilon^{2}_{F}\tau_p}{\tau_s}=\Delta^{2}_{EY}+\left(\frac{4\Delta^{2}_{DP}}{\hbar^{2}}\right)\left(\varepsilon^{2}_{F}\tau^{2}_{p}\right).
\end{equation}
When plotting $\frac{\varepsilon^{2}_{F}\tau_p}{\tau_s}$ versus $\varepsilon^{2}_{F}\tau^{2}_{p}$ for our data in Fig.\ref{fig:Fig4}c we clearly observe a linear behavior for all devices. Using a linear relation we determine the slope and intersect with the y-axis, which directly gives measures for both $\Delta_{EY}$ and $\Delta_{DP}$. The deviating behavior for the p-doped side of device 3 originates from the unexpected relation between $\tau_s$ and n (Fig.\ref{fig:Fig3}b) and $\Delta_{EY}$ could not be extracted. Looking at the respective relaxation rates, we find that tuning the charge carrier density in fact leads to very similar rates for both the DP and EY contribution. For DP we find rates in the range from $1\times10^{9}$~s$^{-1}$ to $2\times10^{10}$~s$^{-1}$ and for EY from $3\times10^{8}$~s$^{-1}$ to $4\times10^{10}$~s$^{-1}$. 

For comparison the analysis using Eq.\ref{eq:1} is applied to the room temperature spin transport data achieved for a SiO$_2$ based device in Ref.~\cite{JozsaLinScale}. In this case we find $\Delta_{DP}\approx90$~$\mu$eV and $\Delta_{EY}\approx2.3$~meV, which is very similar to the result for the h-BN based devices. Likewise, the relaxation rates for both mechanisms are comparable as well, with values in the order of $10^{9}$~s$^{-1}$ for both mechanisms. 

We measured spin transport in supported graphene devices at higher mobilities then achieved so far. Interestingly, we do not see an effect on $\tau_{s}$, which is consistent with results that were recently achieved for SiO$_2$ based devices with tunable mobility~\cite{HanMuDep}. The use of a h-BN substrate allowed us to exclude the influence of the SiO$_2$ substrate and graphene roughness on spin relaxation. These two types of graphene devices only have the contacts and the contaminants in common which therefore appear to dominate the spin relaxation properties of graphene devices.
The contacts are known to have an effect~\cite{HanTunnelCts}, but the mismatch between experimental observations and theoretical predictions is not accounted for. We observe an enhanced spin relaxation time when using TiO$_2$ barriers, which are more smooth than Al$_2$O$_3$. The resistance of our contacts is expected to be sufficient for accurate measurement of non-local spin signals~\cite{PopinciucR}, so the contacts are excluded as the dominant limiting factor for spin relaxation. Other factors that can be held responsible are scattering by covalently bonded adsorbates and charged impurities~\cite{CastroNetoImp}. An interesting approach for future experiments would be to realize clean graphene flakes by removing resist residues and other contaminants. First steps have been taken by the realization of suspended graphene flakes that can be current annealed~\cite{GuimaraesSuspSpin}. For h-BN based devices alternative cleaning methods, as for example mechanical cleaning~\cite{GoossensAFM}, may be adopted to preserve the contacts.

In conclusion we have fabricated graphene spin transport devices based on h-BN crystals with mobilities up to 40~000~cm${^2}$V$^{-1}$s$^{-1}$ that show superior spin relaxation lengths at room temperature. This is directly demonstrated by spin-valve measurements over distances up to 20~$\mu$m and confirmed by Hanle spin precession measurements, which reveal spin relaxation lengths up to 4.5~$\mu$m. The increase with respect to lower mobility SiO$_2$ based devices is due to an increase in the diffusion constant, the spin relaxation time remains virtually unchanged. This is an experimental indication that the substrate and roughness are not the limiting factors for spin relaxation in graphene, which is an important observation since most research has been done on SiO$_2$ based devices. Other factors may be dominant, such as contaminants on the graphene flake. Concerning the relaxation mechanism, we find that our data is best described by a combination of the EY and DP mechanisms. The respective relaxation rates are found to be very similar for both mechanisms, indicating that the spin relaxation is not dominated by a single mechanism.

We acknowledge B. Wolfs, J. G. Holstein and H. M. de Roosz for their technical assistance. 
This work is financially supported by the Dutch Foundation for Fundamental Research on Matter (FOM), NWO, NanoNed and the Zernike Institute for Advanced Materials.


\providecommand{\noopsort}[1]{}\providecommand{\singleletter}[1]{#1}%

\end{document}